\documentclass{article}
\usepackage{multirow}
\usepackage[preprint]{neurips_2026}
\usepackage{subcaption}
\usepackage{booktabs}
\usepackage[utf8]{inputenc}
\usepackage[T1]{fontenc}
\usepackage{graphicx}
\usepackage{hyperref}
\usepackage{longtable}
\usepackage{parskip}
\usepackage{array}
\usepackage{caption}
\usepackage{url}
\usepackage[table]{xcolor}
\usepackage{xcolor}        
\definecolor{lightmud}{RGB}{230, 220, 200}
\usepackage{longtable} 
\usepackage{url}
\usepackage[utf8]{inputenc} 
\usepackage[T1]{fontenc}    
\usepackage{hyperref}       
\usepackage{url}            
\usepackage{booktabs}       
\usepackage{amsfonts}       
\usepackage{nicefrac}       
\usepackage{microtype}      
\usepackage{caption}
\usepackage{url}
\usepackage[table]{xcolor} 
\usepackage{xcolor}         
\definecolor{lightmud}{RGB}{230, 220, 200} 
\usepackage{longtable}
\usepackage{hyperref}
\usepackage{xcolor}
\definecolor{annotRed}{RGB}{220,50,47}
\definecolor{annotPink}{RGB}{213,94,170}
\definecolor{annotTeal}{RGB}{0,150,150}
\usepackage{pifont}
\usepackage{makecell}

\title{DentiAsk: A VQA Benchmark for Multimodal Reasoning in Panoramic Dental Radiographs}

\author{
\normalfont\mdseries
Debesh Jha\textsuperscript{1,*},
Tapas Kumar Dutta\textsuperscript{2,*},
Roshan Paudel\textsuperscript{1},
Onkar Kishor Susladkar\textsuperscript{3},
Aashish Ghimire\textsuperscript{1},\\
Anupam Dhakal\textsuperscript{1},
Sabin Adhikari\textsuperscript{1},
Nand Kumar Yadav\textsuperscript{7},
Deepak Ranjan Nayak\textsuperscript{4},\\
Pravin Pawar\textsuperscript{5},
William C. W. Chen\textsuperscript{6},
Glenda Reynolds\textsuperscript{7}\\[1mm]
\textsuperscript{1}Biomedical Perception \& Intelligence Lab, University of South Dakota, USA\\
\textsuperscript{2}University of Surrey, United Kingdom\\
\textsuperscript{3}University of Illinois Urbana-Champaign, USA\\
\textsuperscript{4}Malaviya National Institute of Technology Jaipur, India\\
\textsuperscript{5}CSMSS Dental College and Hospital, India\\
\textsuperscript{6}Biomedical \& Translational Sciences, Sanford School of Medicine, University of South Dakota, USA\\
\textsuperscript{7}Department of Dental Hygiene, University of South Dakota, USA\\
\texttt{debesh.jha@usd.edu}\\
\textsuperscript{*}Joint first authors.
}

\begin{document}

\maketitle

\begin{abstract}
Accurate interpretation of panoramic dental radiographs requires the integration of multiple reasoning capabilities: detection, spatial localization, and quantitative assessment. Despite recent advances in multimodal learning, existing medical visual question answering (VQA) benchmarks do not fully capture this complexity, often reducing the task to simplified classification or templated queries. As a result, they provide limited coverage of the diverse reasoning processes required for clinically meaningful interpretation. We introduce DentiAsk, a large-scale dental VQA benchmark that pairs high-resolution panoramic dental radiographs with clinician-validated question–answer pairs spanning three reasoning tiers: i.e., descriptive, spatial localization, and numerical quantification across three high-prevalence pathologies: periapical radiolucency (PARL), impacted teeth, and dental caries. DentiAsk comprises 1,000 high-resolution radiographs annotated with 10,000 expert curated QA pairs. To our knowledge, it is the first dental VQA benchmark to unify categorical, spatial, and quantitative reasoning as separately scored tasks within a single evaluation framework. We benchmark 10 state-of-the-art vision–language models (LLaVA-v1.5, LLaVA-v1.6, Qwen-VL, InternVL2, and LLaVA-Med) and find that models achieve stronger performance on descriptive queries, whereas models degrade sharply on spatial localization and counting, exposing limitations in compositional, multi-step reasoning. These findings reveal a gap between visual recognition and clinically meaningful reasoning, establishing DentiAsk as a challenging benchmark for advancing multimodal reasoning in medical imaging. Dataset: \url{https://osf.io/sd7fk/overview}; and code \url{https://github.com/DebeshJha/DentiAsk}.

\end{abstract}

\section{Introduction}

\begin{figure}
    \centering
    \includegraphics[width=0.99\linewidth, trim={0 0 0 0}, clip]{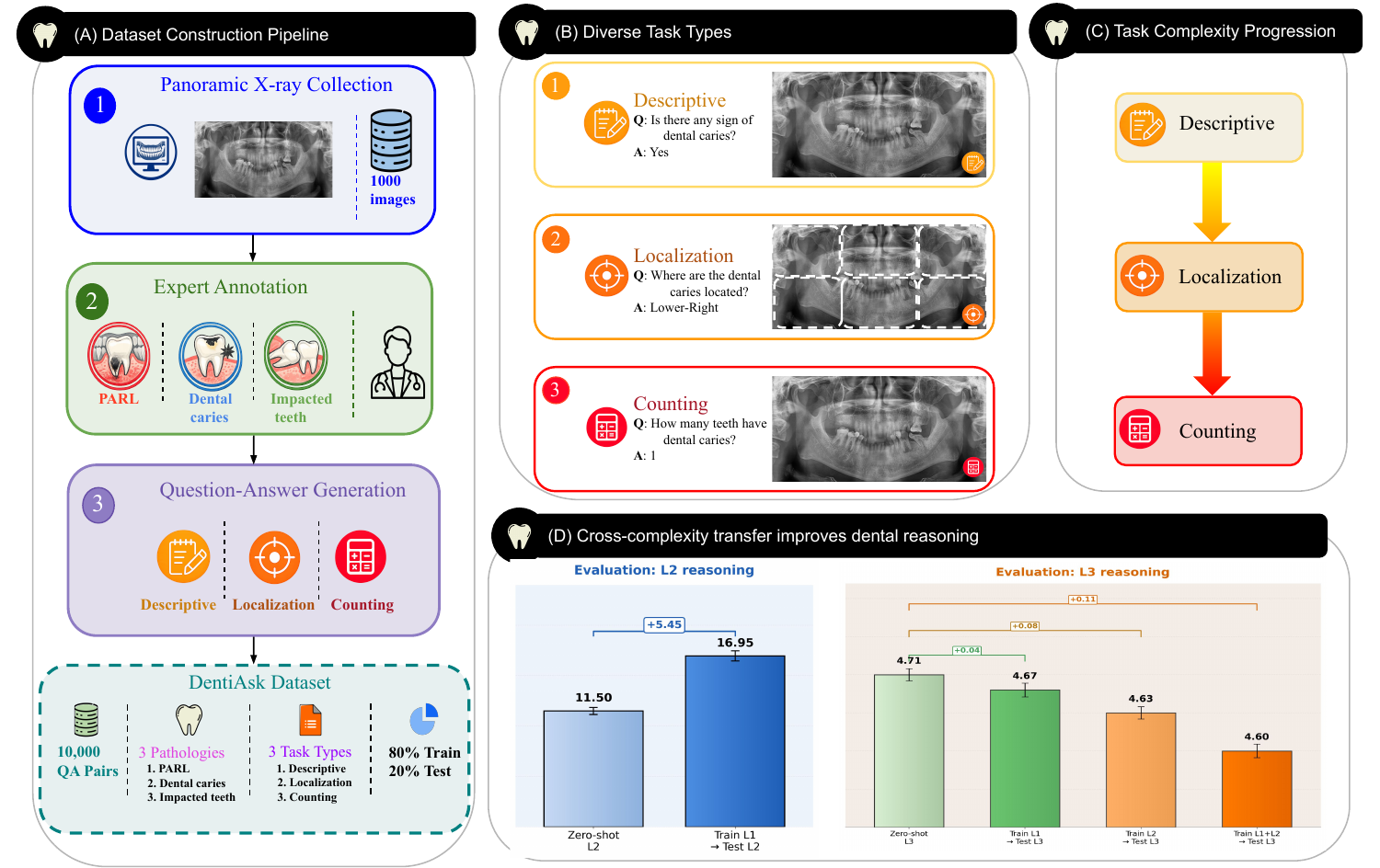}
    \caption{Overview of DentiAsk. (A) Construction pipeline: 1,000 panoramic radiographs are expert annotated for three pathologies (PARL, dental caries, impacted teeth) and converted into 10,000 image-question instances spanning three task types, split 80/20 into train/test. (B) Task types per image: descriptive (L1), localization (L2), and counting (L3). (C) Complexity progression: descriptive $\rightarrow$ localization $\rightarrow$ counting. (D) Cross-complexity transfer: training on lower tiers improves performance on harder, held-out tiers over zero-shot (L2 reported as Subset Accuracy; L3 as RMSE), with the largest L3 gain from combined L1+L2. }
    
    \label{fig:Architectural_design}
\end{figure}

Panoramic dental radiographs are a first line imaging modality in clinical
dentistry, and the conditions they reveal periapical radiolucency (PARL), impacted teeth, and dental caries are among the most prevalent oral health problems worldwide~\cite{peres2019}. Interpreting these images is not a single act of recognition. A clinician's reasoning is query driven: depending on the diagnostic question, the same radiograph must be read to detect whether a condition is present, to localize it within the dental arch, or to count how many teeth are affected. A single panoramic view often contains co-occurring conditions, overlapping anatomical structures, and missing or partially visible teeth, so accurate interpretation depends on flexibly switching between these distinct reasoning modes within one image.

Multimodal large language models (MLLMs) have advanced rapidly on joint
visual–textual reasoning~\cite{alayrac2022flamingo, liu2023llava, dai2023instructblip}, and medical visual question answering (VQA) benchmarks such as VQA-RAD~\cite{lau2018vqarad}, PathVQA~\cite{he2020pathvqa}, and
SLAKE~\cite{liu2021slake} have extended this paradigm to clinical imaging. Yet these benchmarks largely evaluate isolated or simplified tasks, and even recent
large-scale efforts~\cite{zhang2023pmcvqa, hu2024omnimedvqa} remain centered on
single-task evaluation. In dentistry, most public datasets target vision-only
perception—tooth segmentation and pathosis detection (Table~\ref{tab:related_datasets}), while recent dental vision–language resources such as DentalGPT~\cite{cai2025dentalgpt}, MMOral~\cite{hao2025mmoral}, and DentVLM~\cite{meng2025dentvlm} provide large,
open-ended instruction corpora aimed at training generalist models. What remains missing is a controlled benchmark that separates the distinct reasoning operations a clinician performs and scores each one explicitly, so that a model's recognition ability can be distinguished from its spatial and quantitative reasoning.

To address this gap, we introduce \textit{DentiAsk}, a panoramic dental VQA benchmark that pairs 1,000 high-resolution radiographs with 10,000 clinician grounded question–answer pairs (Fig.~\ref{fig:Architectural_design}). DentiAsk organizes diagnostic questions into three reasoning tiers: descriptive recognition, region level localization, and numerical counting across three high-prevalence conditions: PARL, impacted teeth, and dental caries. To keep spatial annotation robust to missing or partially visible teeth, localization uses a six region scheme over the dental arch rather than tooth level indexing. All answers are closed-form (binary or subtype labels, region sets, and integer counts), so every prediction $A = f_{\theta}(I, Q)$ for a radiograph $I$ and query $Q$ is directly verifiable. The questions are instantiated from clinician-defined templates over expert segmentation annotations, making the benchmark both clinically grounded and reproducible.

We benchmark ten state-of-the-art VLMs on DentiAsk and observe a consistent
perception–reasoning gap: models answer recognition queries well but degrade sharply on localization and counting, with even the strongest model reaching only 1.4 MAE. A cross-complexity analysis further shows that reasoning ability transfers from simpler to harder tiers but does not close this gap. Together, these results position DentiAsk as a challenging, diagnostically grounded testbed for multimodal reasoning in dental radiology.

\noindent The main contributions of this work are as follows:
\begin{enumerate}
    \item \textbf{A panoramic dental VQA benchmark.} We introduce \textit{DentiAsk}, 1,000 panoramic radiographs paired with 10,000 closed-form question–answer pairs over three high-prevalence pathologies (PARL, impacted teeth, and dental caries), derived from expert segmentation annotations.

    \item \textbf{A tiered, closed-form evaluation framework.} DentiAsk separates descriptive recognition, region-level localization (via a six-region scheme), and numerical counting as separately scored tasks over the same image, enabling per-tier analysis.

    \item \textbf{Large-scale benchmarking.} We evaluate ten
    state-of-the-art vision–language models under image-level and cross-complexity protocols for the best model.

    \item \textbf{Evidence of a perception–reasoning gap.} Recognition is comparatively tractable, whereas localization and counting remain difficult (best counting MAE $1.4$) and are not closed by model scale or by transfer from simpler tasks, exposing a key limitation of current multimodal systems.
\end{enumerate}

\section{Related Works}

\begin{table}[t]
\centering
\footnotesize
\setlength{\tabcolsep}{4pt}
\begin{tabular}{l c r c c}
\toprule
\textbf{Dataset} & \textbf{Year} & \textbf{Human-verified VQA} &
\textbf{Reasoning Structure} &
\textbf{ROI} \\
\midrule

\multicolumn{5}{l}{\textit{\textcolor{gray}{Dental vision-language datasets}}} \\
\hline
DENT\_VQA~\cite{mbarek2025methodological}  (single-task)   & 2025 & 12,000+   & Single-tier & \ding{55}  \\
DentalGPT~\cite{cai2025dentalgpt} (open-ended)    & 2025 & - & Single-tier & \ding{55} \\
MMOral~\cite{hao2025mmoral} (multi-format)     & 2025 & 1100  & Single-tier & \ding{55} \\
DentVLM~\cite{meng2025dentvlm} (generalist)      & 2025 & 3105 & Single-tier & \ding{55} \\
\hline

\multicolumn{5}{l}{\textit{\textcolor{blue}{Multi-tier diagnostic benchmark}}} \\
\hline
\textbf{DentiAsk (ours)} &
\textbf{2026} &
\textbf{10,000} &
\textbf{Three-tier (L1--L3)} &
\textbf{\checkmark} \\
\bottomrule

\end{tabular}

\caption{Comparison of dental vision-language datasets. DentiAsk is the first benchmark
to explicitly separate diagnostic reasoning into three independently evaluated tiers:
descriptive (L1), localization L2), and counting-based
quantification (L3). } 
\vspace{2pt}
\raggedright
\footnotesize
\label{tab:related_datasets}
\end{table}

Vision Language Models (VLMs) jointly learn visual and textual representations and have become a dominant paradigm for multimodal learning, enabling tasks such as image captioning, visual question answering (VQA), and report generation.  Contrastive pretraining, introduced by CLIP~\cite{radford2021learning}, enabled strong zero-shot transfer, and instruction tuned models such as LLaVA~\cite{liu2023llava}, Qwen-VL~\cite{Bai2023QwenVLAV}, and InternVL~\cite{chen2024internvl} extended these capabilities to general purpose multimodal reasoning. In the medical domain, VLMs have been increasingly explored for clinical reasoning tasks. Datasets such as VQA-RAD~\cite{lau2018vqarad} and PathVQA~\cite{he2020pathvqa} provide medical images paired with question-answer annotations, enabling supervised training for medical VQA. More recent datasets such as Kvasir-VQA~\cite{gautam2025kvasir} extend this paradigm with larger-scale and more diverse annotations.  Domain adapted models, including Med-Flamingo~\cite{moor2023med} and the spatially grounded PRS-Med~\cite{trinh2026prs}, further demonstrate the value of combining visual, textual, and spatial supervision for clinically relevant
reasoning.

Dental imaging research, by contrast, has until recently centered on vision only perception. As summarized in Table~\ref{tab:related_datasets}, DENT\_VQA reports 12K+ expert-formulated QA pairs; DentalGPT verifies only image labels, with QA pairs being LLM generated. MMOral curates 1,100 QA pairs (500 closed + 600 open) via manual quality checks rather than documented per-pair dentist review, while DentVLM's 3,105 QA pairs are explicitly ground truth labeled by four expert dentists with confidence filtering and voting, the most rigorously verified of the four. Prior dental datasets target segmentation or detection rather than language-grounded reasoning: 3DTeethSeg'22~\cite{ben20233dteethseg} for intraoral 3D tooth segmentation, DENTEX~\cite{hamamci2023dentex} for tooth enumeration and pathosis detection on panoramic X-rays, the pediatric caries dataset of Zhang et al.~\cite{zhang2023children}, the multimodal CBCT/panoramic collection of Huang et al.~\cite{huang2024multimodal}, and STS-Tooth~\cite{wang2025multi} for semi-supervised segmentation. These resources advance dental perception but provide no aligned question–answer supervision, leaving clinically grounded diagnostic reasoning largely unevaluated.

There are several recent efforts in dental imaging, as shown in Table~\ref{tab:related_datasets}. DentalGPT~\cite{cai2025dentalgpt} aligns radiographic features with textual descriptions and QA pairs and improves reasoning through reinforcement learning; MMOral~\cite{hao2025mmoral} and DentVLM~\cite{meng2025dentvlm} contribute large instruction-following corpora covering a broad range of diagnostic tasks; and DENT\_VQA~\cite{mbarek2025methodological} provides panoramic QA across several diagnostic themes. These datasets primarily target broad, open-ended instruction following and generalist model training. DentiAsk takes a complementary, evaluation-focused stance: it isolates three reasoning tiers—descriptive recognition, region level localization, and explicit counting—as separately scored, closed-form tasks over the same radiograph, across three high-prevalence pathologies (PARL, impacted teeth, and dental caries). To our knowledge, no publicly available dataset evaluates these three reasoning tiers as distinct closed-form tasks over PARL, impacted teeth, and dental caries within a single panoramic radiograph, which lets model performance be analyzed per tier.

\section{The DentiAsk dataset }
\subsection*{Design goal}
\begin{figure}[t!]
    \centering
    \includegraphics[width=1\textwidth, height=0.15\textheight, trim=0 0 0 0, clip]{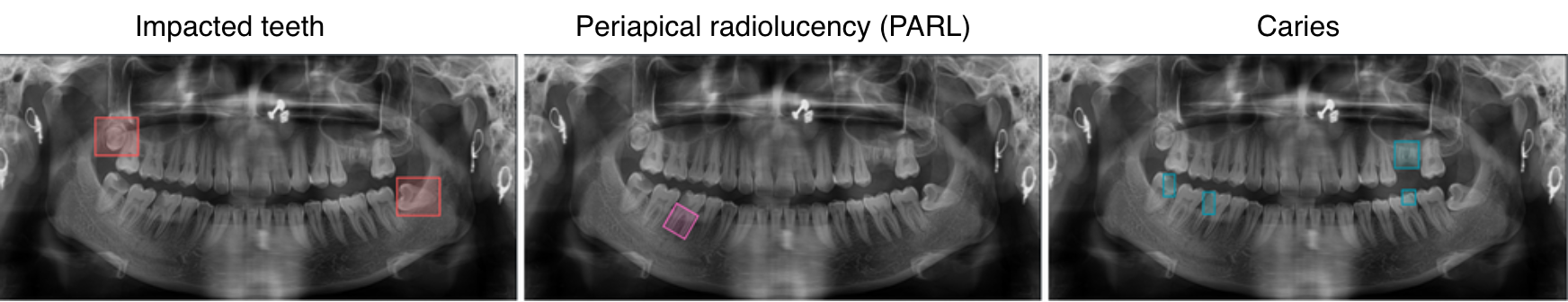}
    \caption{Three target dental conditions in DentiAsk: impacted teeth, periapical radiolucency (PARL), and dental caries, shown from left to right. Colored overlays indicate expert clinician annotations used to derive the diagnostic question--answer pairs: impacted teeth in \textcolor{annotRed}{red}, PARL in \textcolor{annotPink}{pink}, and dental caries in \textcolor{annotTeal}{teal}.}
    \label{fig:dental_xraythree}
\end{figure}

We construct DentiAsk to evaluate whether multimodal models can perform query-conditioned diagnostic reasoning on panoramic dental radiographs, rather than image level recognition alone. A single panoramic view may contain overlapping anatomical structures, such as missing or partially visible teeth, multiple co-existing pathologies, or subtle radiological findings, so correct interpretation requires more than detecting the presence or absence of disease. To capture this, DentiAsk targets three clinically common conditions that each demand a different reasoning skill (Fig.~\ref{fig:dental_xraythree}): periapical radiolucency (PARL), which requires subtle lesion recognition near the root apices; impacted teeth, which require spatial and anatomical interpretation; and dental caries, which require detecting localized structural loss.

\subsection*{Data collection}
Panoramic radiographs were retrospectively acquired from Chhatrapati Shahu Maharaj Shikshan Sanstha Dental College and Hospital, India using two clinical Panoramic radiography units, namely the Planmeca ProMax 2D and Sirona Orthophos XG. All the images were anonymized at the source before transfer to the research team.  The dataset is released for non-commercial research use only. Ambiguous pediatric cases involving uncertain tooth development were excluded after consultation with experienced dental practitioners, since classifying such teeth as impacted could introduce unreliable supervision.

\subsection*{Annotation protocol}
An experienced licensed dentist annotated the impacted teeth, PARL, and caries on each panoramic radiograph using LabelBox~\cite{labelbox2024}. We exported the labels as a .json file and converted them into structured ground truth. This ground truth served as the foundation for our VQA-pair construction. Using segmentation annotations from DentiMap~\cite{jha2026dentimap}, we constructed VQA pairs based on 10 predefined diagnostic question templates (Table~\ref{tab:original}). The templates were designed in collaboration with dental experts to cover disease recognition, localization, subtype classification, and numerical counting tasks. Four trained annotators generated question--answer pairs from the original radiographs and their corresponding ground-truth annotations. Each generated pair was checked for consistency with the visual findings, disease labels, region annotations, and numerical counts.

\begin{figure}[!b]
    \centering
      \includegraphics[width=0.7\textwidth, trim=0 0 0 0, clip]{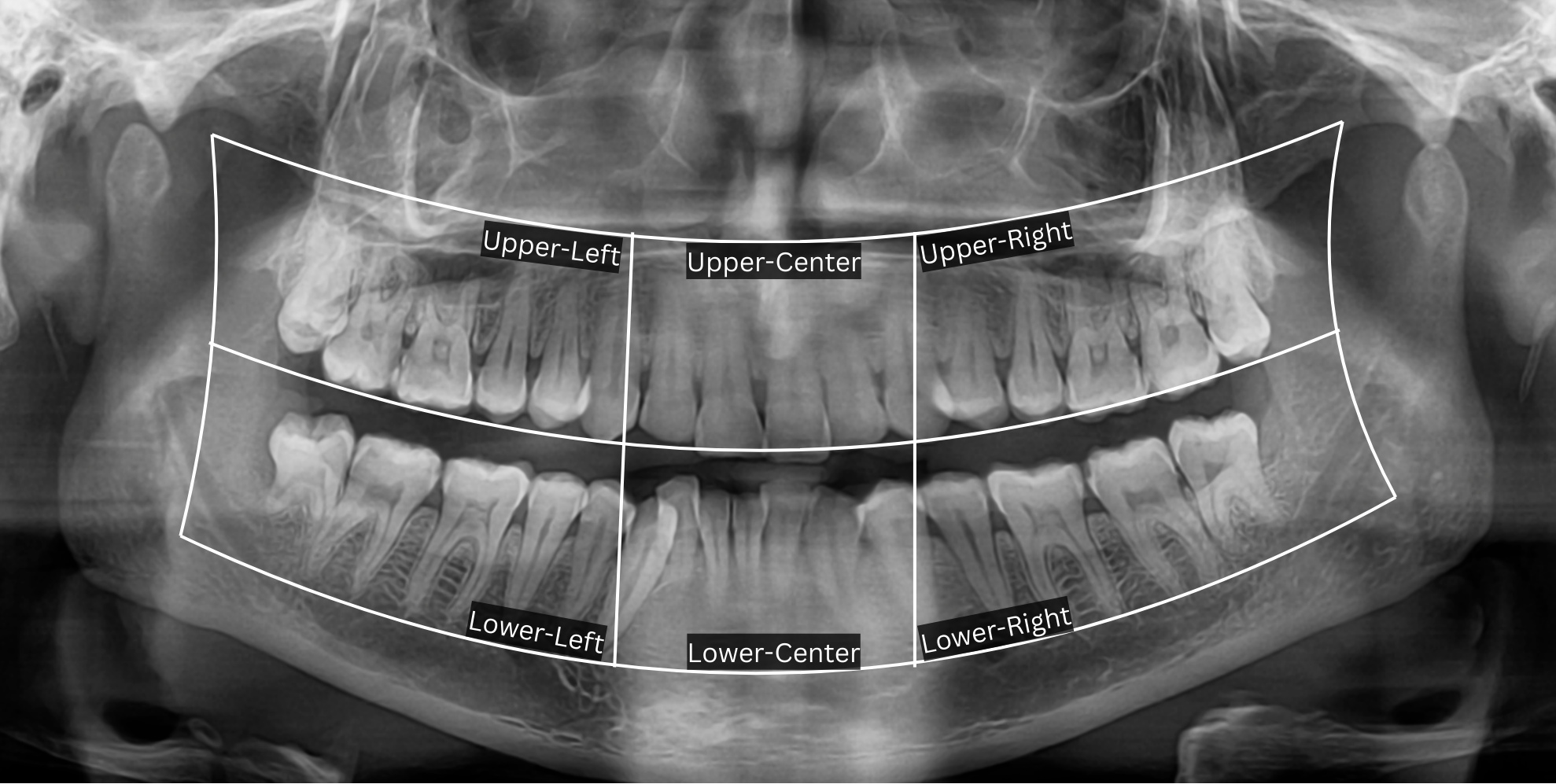}
    \caption{Demonstration of the six-region grid employed for localization tasks. Since tooth level numbering is unreliable on panoramic radiographs with missing or partially visible teeth, we divide the dental arch into six regions: Upper-Left, Upper-Center, Upper-Right, Lower-Left, Lower-Center, and Lower-Right. Each finding is assigned to every region it overlaps, so a localization answer can include more than one region.}
    \label{fig:dental_xray1}
\end{figure}

\subsection*{Region-based spatial localization} 
A key challenge in panoramic dental VQA is spatial annotation. Although tooth level numbering systems such as the Universal Numbering System are widely used in clinical practice, direct tooth indexing is unreliable for benchmark construction because panoramic radiographs often exhibit missing teeth, incomplete dentition, developmental variation, or partial visibility. To address this limitation, we use a region-based localization framework that divides the dental arch into six anatomical zones: \textit{Upper-Left, Upper-Center, Upper-Right, Lower-Left, Lower-Center, and Lower-Right}. Fig. ~\ref{fig:dental_xray1} demonstrates the six-region grid employed for localization tasks.  This six-region strategy provides a clinically interpretable yet robust spatial representation. It preserves meaningful anatomical localization while reducing annotation ambiguity caused by tooth absence or variable image quality. As a result, DentiAsk supports spatial reasoning in a way that is both clinically practical and computationally consistent.

\begin{table*}[!t]
\centering
\setlength{\tabcolsep}{5pt}
\resizebox{\textwidth}{!}{
\begin{tabular}{|c|p{2.8cm}|p{5.2cm}|p{6.2cm}|}
\hline
\rowcolor{lightmud}
\textbf{No} & \textbf{Type} & \textbf{Question} & \textbf{Answer options} \\
\hline

\rowcolor{lightmud}
\multicolumn{4}{|c|}{\textbf{Periapical Radiolucency (PARL)}} \\
\hline

1 & Descriptive & Is there any sign of PARL? & Yes / No \\
\hline

2 & Localization & Where is the PARL located? &
Upper-Left / Upper-Center / Upper-Right / Lower-Left / Lower-Center / Lower-Right / Not Applicable \\
\hline

3 & Numerical Count & How many teeth have PARL? & [0--32] \\
\hline

\rowcolor{lightmud}
\multicolumn{4}{|c|}{\textbf{Impacted Teeth}} \\
\hline

4 & Descriptive & Are there any impacted teeth? & Yes / No \\
\hline

5 & Descriptive & Are the teeth partially impacted, fully impacted, or both? &
Fully Impacted / Partially Impacted / Both / Not Applicable \\
\hline

6 & Localization & Where are the impacted teeth located? &
Upper-Left / Upper-Center / Upper-Right / Lower-Left / Lower-Center / Lower-Right / Not Applicable\\
\hline

7 & Numerical Count & How many impacted teeth are present? & [0--32] \\
\hline

\rowcolor{lightmud}
\multicolumn{4}{|c|}{\textbf{Dental Caries}} \\
\hline

8 & Descriptive & Is there any sign of dental caries? & Yes / No \\
\hline

9 & Localization & Where are the dental caries located? &
Upper-Left / Upper-Center / Upper-Right / Lower-Left / Lower-Center / Lower-Right / Not Applicable\\
\hline

10 & Numerical Count & How many teeth have dental caries? & [0--32] \\
\hline

\end{tabular}}

\caption{10 question templates in DentiAsk, covering three pathologies and three reasoning types, each with a fixed set of
answers.}

\label{tab:original}
\end{table*}

\subsection*{Task and question design} 
Each radiograph is paired with 10 diagnostic questions (Table~\ref{tab:original}) spanning three reasoning tiers: descriptive recognition (L1), localization (L2), and counting (L3). The answer
spaces are closed-form and task-specific: descriptive questions take a binary
Yes/No answer, except the impacted-teeth subtype question, whose answer is one of \{Fully Impacted, Partially Impacted, Both, Not Applicable\}; localization answers select one or more of the six regions (or Not Applicable); and counting answers are integers in $[0,32]$. This closed-form design makes every answer directly verifiable.

\subsection*{Final Dataset}
The final dataset consists of 1,000 high-resolution panoramic dental radiographs (each of $2670 \times 1300$ pixels) paired with 10,000 question-answer pairs from the 10 templates.  All images are stored in BMP format to avoid compression artifacts and preserve diagnostic detail. Because a single radiograph may exhibit more than one condition, the per-pathology image counts overlap: 487 radiographs contain PARL, 367 contain impacted teeth, and 876 contain dental caries.

\section{Experimental Setup}

We evaluate DentiAsk as a benchmark for VQA on panoramic radiographs. Given an input panoramic X-ray and a diagnostic question, the model is required to generate an answer. The answer should include binary responses, anatomical region labels, subtype categories, and numerical counts. This helps us to evaluate whether vision–language models can move beyond image level recognition tasks and perform query-conditioned diagnostic reasoning. 

All experiments were implemented in Python using the PyTorch library. The dataset was split with different strategies to explore generalization capabilities across images and question complexity, while also assessing performance improvements obtained through region grounding.
For each splitting strategy, the input images were normalized and data augmentation was performed during training to increase the dataset size and minimize model overfitting. The data augmentation used was brightness and contrast augmentation by 5\% each with 0.5 probability.

\textbf{Stage 1: Generalization across images: }
To explore the model's generalization capabilities across different images, we split the dataset into train (80\% corresponding to 800 images and 8000 VQA pairs) and evaluate the model's performance in test set (20\% corresponding to 200 images and 2000 VQA pairs).
\begin{figure*}[!t]
        \centering
        \includegraphics[width=0.8\linewidth, trim=0 180 80 0, clip]{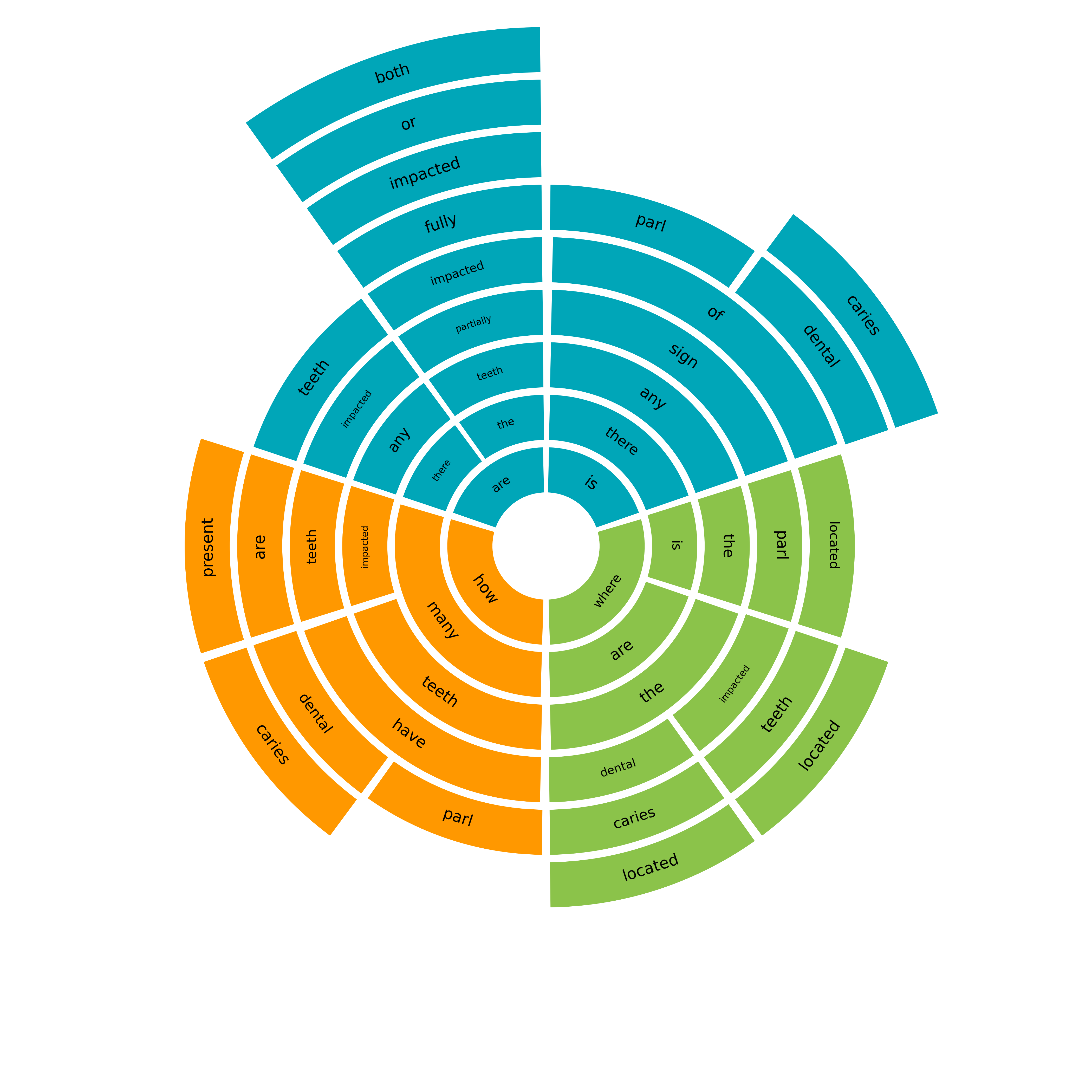}
        \caption{Taxonomy of the 10 DentiAsk diagnostic question templates (T1–T10; Table~\ref{tab:original}). The sunburst radiates from shared question stems to specific templates spanning periapical radiolucency (PARL), impacted teeth, and dental caries. Segments are colored by reasoning complexity: Descriptive (L1, \textcolor{cyan}{teal}), Localization (L2, \textcolor{green}{green}), and Counting (L3, \textcolor{orange}{orange}). } 
        \label{fig:vqa_distribution}
\end{figure*}

\textbf{Stage 2: Generalization across complexity:}
DentiAsk contains 10 questions per image with progressively increasing difficulty levels, comprising 4 descriptive questions (L1), 3 localization questions (L2), and 3 counting questions (L3). This hierarchical design enables the evaluation of a model’s ability to generalize across progressively complex stages of diagnostic reasoning. As illustrated in Fig.~\ref{fig:vqa_distribution}, the benchmark organizes question templates into three reasoning tiers: L1 descriptive recognition questions shown in \textcolor{cyan}{teal}, L2 spatial localization questions shown in \textcolor{green}{green}, and L3 numerical counting questions shown in \textcolor{orange}{orange}. Specifically, we evaluate whether models trained on lower-complexity reasoning stages can generalize to more challenging settings, including training on L1 and evaluating on L2 as reported in Table \ref{tab:cross_L2}, as well as training on L1, L2 and evaluating on L3 as reported in Table~\ref{tab:cross_L3}. For both the experiments, the dataset was first split at the image level using an 80:20 train:test partition. During training, only the L1, L2, or combined L1+L2 question-answer pairs associated with the training images were used, while evaluation was performed on the held-out 20\% test images using their corresponding L2 and L3 question-answer pairs. This protocol ensures that models are evaluated not only on unseen reasoning complexities but also on entirely unseen images, thereby preventing information leakage across reasoning stages. Additionally, we compare these results against out-of-the-box models to analyze how diagnostic reasoning capabilities transfer across increasing levels of complexity involving PARL, impacted teeth, and dental caries.

Since L1, L2, and L3 require answers in task-specific formats, prompt tuning was employed to explicitly inject the expected response structure during all the aforementioned training and inference. In particular, L1 descriptive questions were augmented with the prompt: \textit{``Answer one of (possible answers) only''}, constraining the model to select responses exclusively from the predefined answer space. L2 localization questions used the prompt: \textit{``
It may appear in one or more regions of the image. List all regions that apply, separated by semicolons, using only these values:  (location\_options). Do not include any other text. Example:
Upper-Left;Upper-Right;Lower-Left
''}, restricting outputs to combinations from the predefined anatomical location set. Similarly, L3 counting questions used the prompt: \textit{``Answer with a number only''}, enforcing strictly numerical outputs. This formulation reduces ambiguity in generated responses and enables more reliable evaluation across reasoning stages.

We select metrics aligned with each tier's answer structure: L1 uses Accuracy, Precision, Recall, Macro-F1 for single-label classification over closed-set Yes/No and subtype answers, while L2 uses micro and macro-F1, Jaccard, and Subset Accuracy to handle multi-label region sets, capturing both partial overlap and exact-match correctness. L3 is treated as regression over integer counts, using MAE for average deviation and RMSE to penalize larger miscounts, since affected-tooth counts are continuous rather than categorical. All models were fine-tuned using Low-Rank Adaptation \cite{hu2022lora} rather than full parameter updates, with base model weights kept frozen and only the injected low-rank adapters trained on the task-specific question–answer pairs.

\section{Results and Discussion}
\subsection{Quantitative results}

Following the Stage 1 image level generalization protocol, Table~\ref{tab:l1_descriptive}, Table~\ref{tab:l2_localization}, and Table~\ref{tab:l3_counting} report model performance on the held-out test set across the three reasoning tiers descriptive (L1), localization (L2), and counting (L3) respectively.
Across the settings, InternVL2-8B~\cite{chen2024internvl} consistently achieves the best performance on most evaluation metrics, indicating stronger alignment between generated answers and clinical reference responses. On L1 descriptive recognition (Table~\ref{tab:l1_descriptive}), InternVL2-8B attains the highest accuracy and macro-F1 (0.7562 / 0.4687), with most models clustering between 0.58-0.76 accuracy. However, macro-F1 scores remain considerably lower (0.31-0.47), indicating that this accuracy is not matched by balanced performance across answer classes, and that recognition-level queries, while more tractable than localization or counting, are not yet reliably solved. On L2 spatial localization (Table~\ref{tab:l2_localization}), performance drops markedly under the stricter Subset Accuracy metric (best: 0.4283) despite moderate micro-F1 scores ($\sim$0.60), underscoring the difficulty of exact multi-region matching. On L3 counting (Table~\ref{tab:l3_counting}), all models exhibit substantial regression error, with InternVL2-8B again performing best (MAE 1.4150, RMSE 2.4348), confirming that numerical quantification remains the most challenging reasoning tier across the benchmark. The metric-dependent reversal on L2 InternVL2-8B leading on F1/Jaccard but LLaVA-v1.6-Vicuna-13B leading on Subset Accuracy reflects the trade-off between exact-match and overlap-based multi-label evaluation.

\begin{table}[t]
\centering
\small
\setlength{\tabcolsep}{4pt}
\begin{tabular}{lcccc}
\toprule
\textbf{Model} & \textbf{Acc}$\uparrow$ & \textbf{Prec}$\uparrow$ & \textbf{Rec}$\uparrow$ & \textbf{M-F1}$\uparrow$ \\
\midrule
InternVL2-1B~\cite{chen2024internvl} & 0.7063 & 0.4435 & 0.4516 & 0.4495 \\
InternVL2-2B~\cite{chen2024internvl} & 0.7187 & 0.4563 & 0.4570 & 0.4520 \\
LLaVA-v1.5-7B~\cite{10655294} & 0.6089 & 0.3095 & 0.3387 & 0.3145 \\
LLaVA-v1.6-Mistral-7B~\cite{liu2024llavanext} & 0.6428 & 0.3521 & 0.3769 & 0.3631 \\
LLaVA-Med-v1.5-Mistral-7B~\cite{10.5555/3666122.3667362} & 0.5834 & 0.2987 & 0.3234 & 0.3089 \\
InternVL2-8B~\cite{chen2024internvl} & \textbf{0.7562} & \textbf{0.4548} & \textbf{0.4879} & \textbf{0.4687} \\
LLaVA-v1.6-Vicuna-7B~\cite{liu2024llavanext} & 0.6241 & 0.3239 & 0.3532 & 0.3290 \\
Qwen-VL \cite{Bai2023QwenVLAV} & 0.6388 & 0.3220 & 0.3629 & 0.3297\\
LLaVA-v1.5-13B~\cite{10655294} & 0.6534 & 0.3412 & 0.3845 & 0.3481 \\
LLaVA-v1.6-Vicuna-13B~\cite{liu2024llavanext} & 0.6687 & 0.3500 & 0.3991 & 0.3554 \\
\bottomrule
\end{tabular}
\caption{\textbf{Descriptive recognition (L1) performance under a single-label classification formulation.} Each descriptive question is treated as an independent closed-set prediction over a binary (Yes/No) or subtype answer space across three pathologies (PARL, impacted teeth, and dental caries). We report accuracy (Acc), precision (Prec), recall (Rec), and macro-averaged F1 (M-F1). All metrics range in $[0, 1]$; $\uparrow$ denotes higher is better.}
\label{tab:l1_descriptive}
\end{table}

\begin{table}[t]
\centering
\small
\setlength{\tabcolsep}{5pt}
\begin{tabular}{lcccc}
\toprule
\textbf{Model} & \textbf{mi-F1}$\uparrow$ & \textbf{ma-F1}$\uparrow$ & \textbf{Jacc}$\uparrow$ & \textbf{Sub-A}$\uparrow$ \\
\midrule
InternVL2-1B~\cite{chen2024internvl} & 0.5584 & 0.4015 & 0.5087 & 0.3517 \\
InternVL2-2B~\cite{chen2024internvl} & 0.5744 & 0.4138 & 0.5303 & 0.3683 \\
LLaVA-v1.5-7B~\cite{10655294} & 0.5401 & 0.3652 & 0.5068 & 0.3912 \\
LLaVA-v1.6-Mistral-7B~\cite{liu2024llavanext} & 0.5621 & 0.3901 & 0.5289 & 0.4198 \\
LLaVA-Med-v1.5-Mistral-7B~\cite{10.5555/3666122.3667362} & 0.5123 & 0.3456 & 0.4912 & 0.3687 \\
InternVL2-8B~\cite{chen2024internvl} & \textbf{0.5989} & \textbf{0.4318} & \textbf{0.5617} & 0.3817 \\
LLaVA-v1.6-Vicuna-7B~\cite{liu2024llavanext} & 0.5533 & 0.3772 & 0.5220 & 0.4063 \\
Qwen-VL \cite{Bai2023QwenVLAV} & 0.5568 & 0.3782 & 0.5217 & 0.4082\\
LLaVA-v1.5-13B~\cite{10655294} & 0.5687 & 0.3998 & 0.5378 & 0.4156 \\
LLaVA-v1.6-Vicuna-13B~\cite{liu2024llavanext} & 0.5745 & 0.4073 & 0.5445 & \textbf{0.4283} \\
\bottomrule
\end{tabular}
\caption{\textbf{Spatial localization (L2) performance under a multi-label classification formulation.} Each localization question requires predicting one or more of six anatomical regions (\textit{Upper-Left, Upper-Center, Upper-Right, Lower-Left, Lower-Center, Lower-Right}, or \textit{Not Applicable}) per pathology. We report micro-averaged F1 (mi-F1), macro-averaged F1 (ma-F1), Jaccard similarity (Jacc), and subset accuracy (Sub-A), where Sub-A requires an exact match across all predicted region labels and represents the strictest criterion. All metrics range in $[0, 1]$; $\uparrow$ denotes higher is better.}
\label{tab:l2_localization}
\end{table}

\begin{table}[t]
\centering
\small
\setlength{\tabcolsep}{6pt}
\begin{tabular}{lcc}
\toprule
\textbf{Model} & \textbf{MAE}$\downarrow$ & \textbf{RMSE}$\downarrow$ \\
\midrule
InternVL2-1B~\cite{chen2024internvl} & 1.5750 & 2.8679 \\
InternVL2-2B~\cite{chen2024internvl} & 1.5200 & 2.5430 \\
LLaVA-v1.5-7B~\cite{10655294} & 1.6342 & 3.0128 \\
LLaVA-v1.6-Mistral-7B~\cite{liu2024llavanext} & 1.5487 & 2.8934 \\
LLaVA-Med-v1.5-Mistral-7B~\cite{10.5555/3666122.3667362} & 1.7234 & 3.1456 \\
InternVL2-8B~\cite{chen2024internvl} & \textbf{1.4150} & \textbf{2.4348} \\
LLaVA-v1.6-Vicuna-7B~\cite{liu2024llavanext} & 1.5950 & 2.9662 \\
Qwen-VL \cite{Bai2023QwenVLAV} & 1.7129 & 3.1028\\
LLaVA-v1.5-13B~\cite{10655294} & 1.4956 & 2.8012 \\
LLaVA-v1.6-Vicuna-13B~\cite{liu2024llavanext} & 1.4781 & 2.7541 \\
\bottomrule
\end{tabular}
\caption{\textbf{Numerical counting (L3) performance under a regression formulation.} Each counting question requires predicting an integer count of affected teeth per pathology, treated as a continuous regression target over the range $[0, 32]$. We report Mean Absolute Error (MAE) and Root Mean Squared Error (RMSE); $\downarrow$ denotes lower is better.}
\label{tab:l3_counting}
\end{table}

\begin{table}
\centering
\begin{tabular}{cc|cccc}
\hline
\textbf{Training} & \textbf{Evaluation} & \textbf{mi-F1 $\uparrow$} & \textbf{ma-F1}$\uparrow$ & \textbf{Jacc}$\uparrow$  & \textbf{Sub-A}$\uparrow$ \\
\hline
- & L2 & 0.3344 &0.3034 & 0.2509 & 0.1150 \\
L1 & L2 & 0.3811 &0.3410 & 0.2941 & 0.1695 \\
\hline
\end{tabular}
\caption{Cross-complexity generalization analysis for InternVL2-8B~\cite{chen2024internvl} trained on lower-complexity tiers L1 and evaluated on harder, held-out tiers. ``-'' denotes the zero-shot baseline (no task-specific training).}
\label{tab:cross_L2}
\end{table}

\begin{table}
\centering
\begin{tabular}{cc|cc}
\hline
\textbf{Training} & \textbf{Evaluation} & \textbf{MAE $\downarrow$} & \textbf{RMSE}$\downarrow$ \\
\hline
- & L3 & 2.55 & 4.71 \\
L1 & L3 & 2.52 & 4.67 \\
L2 & L3 & 2.50 & 4.63 \\
L1+L2 & L3 & \textbf{2.46} & \textbf{4.60} \\
\hline
\end{tabular}
\caption{Cross-complexity generalization analysis for InternVL2-8B~\cite{chen2024internvl} trained on lower-complexity tiers (L1, L2, or L1+L2) and evaluated on harder, held-out tiers. ``-'' denotes the zero-shot baseline (no task-specific training).}
\label{tab:cross_L3}
\end{table}

\begin{figure}[t!]
    \centering
    \includegraphics[width=\textwidth]{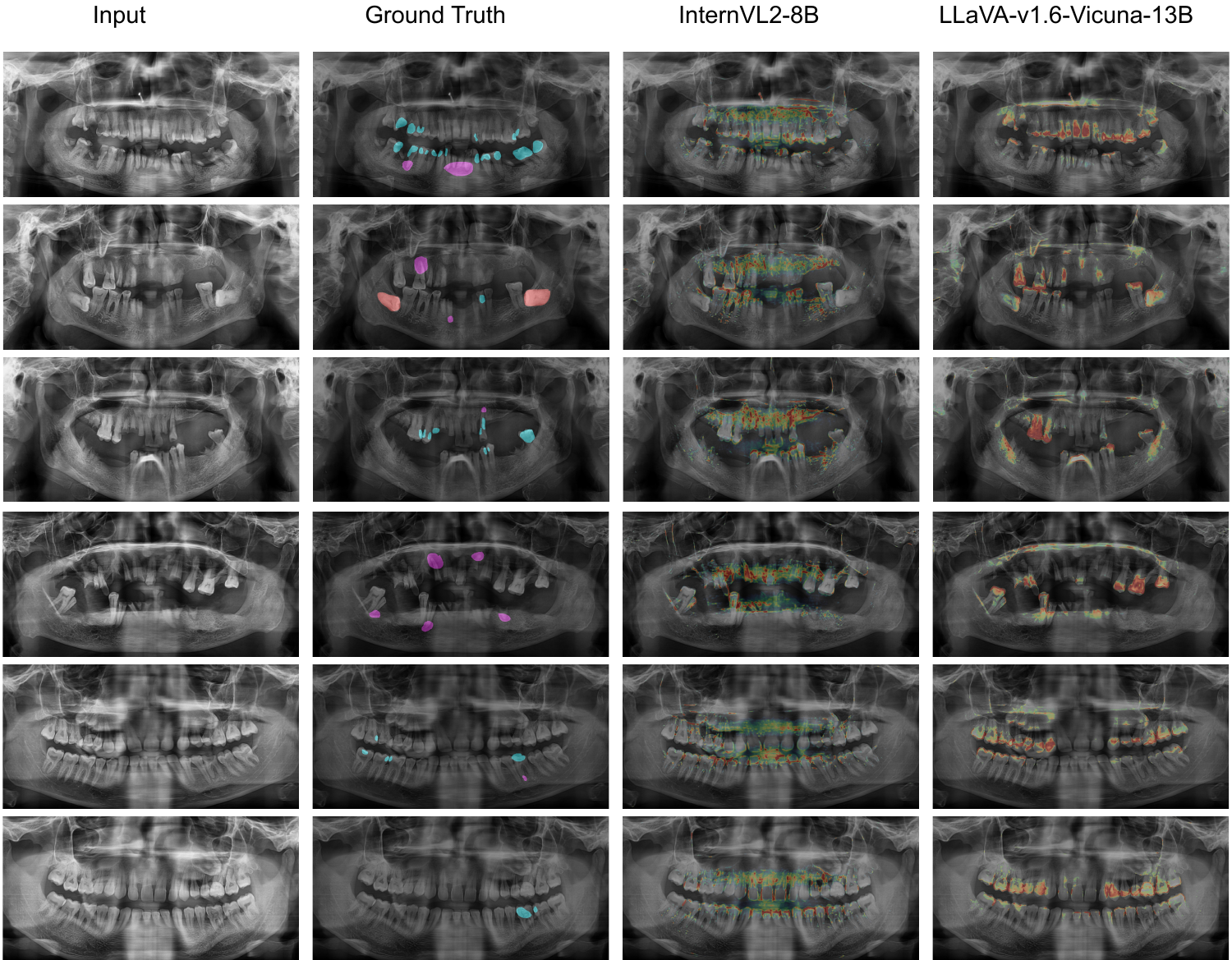}
    \caption{Qualitative comparison of expert-annotation and model-generated attention heatmaps on panoramic dental radiographs. InternVL2-8B~\cite{chen2024internvl} exhibits broader, more diffuse  patterns that capture general regions whereas LLaVA-v1.6-Vicuna-13B~\cite{liu2024llavanext}, produces sharper, more focused but occasionally misses the target pathology entirely.}
    \label{gradcam}
\end{figure}

Tables~\ref{tab:cross_L2} and \ref{tab:cross_L3} examine whether reasoning ability transfers from simpler to harder tiers for InternVL2-8B. For L2 localization (Table~\ref{tab:cross_L2}), training on L1 alone improves all metrics over the zero-shot baseline, with Subset Accuracy rising from 0.1150 to 0.1695 and Jaccard from 0.2509 to 0.2941, indicating that descriptive supervision provides transferable spatial cues. For L3 counting (Table~\ref{tab:cross_L3}), error decreases progressively as more lower-tier supervision is added, with MAE dropping from 2.55 (zero-shot) to 2.52 (L1) and 2.50 (L2), reaching its lowest value of 2.46 when L1 and L2 are combined, mirroring the RMSE trend (4.71 $\rightarrow$ 4.60). These results confirm that cross-complexity transfer is cumulative but only partially closes the gap to in-tier supervision.

\subsection{Qualitative results}
Fig. ~\ref{gradcam} presents qualitative Grad-CAM visualizations of the top 10\% most activated regions for LLaVA-v1.6-Vicuna-13B~\cite{liu2024llavanext} and Intern-VL2-8B~\cite{chen2024internvl}, alongside the input panoramic radiographs and expert-annotated segmentation masks. Although neither model is explicitly trained for segmentation, both exhibit attention overlap with clinically relevant pathological regions, indicating the emergence of anatomically grounded representations through VQA supervision alone. However, notable differences in localization behavior are observed between the models. Intern-VL2-8B produces broader and more spatially diffuse patterns that span larger portions of the dentition and surrounding anatomical structures, yet consistently covers the correct pathological regions. Conversely, LLaVA-v1.6-Vicuna-13B demonstrates comparatively sharper and more localized heatmaps; however, this high focus occasionally causes it to miss the true clinical ground truth entirely by focusing on unrelated localized structures. These results suggest that reasoning oriented supervision can implicitly induce clinically meaningful visual grounding, while also highlighting differences in localization precision and contextual reliance across foundation VLMs. Overall, these results suggest that reasoning capabilities in dental VQA emerge progressively, where learning lower level descriptive and localization knowledge provides a strong foundation for handling more complex counting and diagnostic reasoning tasks.

These results highlight two key observations. First, existing vision-language models remain limited in their ability to perform question-specific reasoning within medical imaging contexts, particularly for clinically grounded diagnostic interpretation. Second, dataset composition, especially the inclusion of multiple question–answer pairs per image, significantly influences task complexity and reasoning difficulty. Collectively, these findings position \textit{DentiAsk} as a challenging benchmark that exposes important limitations of current models in clinically relevant visual understanding and diagnostic reasoning. 

\subsection{Discussion and Limitations}

Tables~\ref{tab:l1_descriptive} and \ref{tab:l3_counting} show a clear, graded gap across reasoning tiers: descriptive recognition is the most tractable (though the accuracy–macro-F1 gap in Table 3 suggests this reflects uneven class-wise performance rather than reliable detection), while localization and counting are markedly harder. InternVL2-8B is the strongest model overall, achieving the best accuracy and macro-F1 on L1 (0.7562 / 0.4687, Table~\ref{tab:l1_descriptive}), the best micro-F1, macro-F1, and Jaccard on L2 (0.5989 / 0.4318 / 0.5617, Table~\ref{tab:l2_localization}), and the lowest MAE/RMSE on L3 (1.4150 / 2.4348, Table~\ref{tab:l3_counting})---though LLaVA-v1.6-Vicuna-13B attains the highest Subset Accuracy on L2 (0.4283), the strictest exact-match criterion. A plausible reading is that descriptive answering is supported by coarse, image level alignment between visual and textual features, whereas counting requires instance level enumeration and localization requires binding a finding to a specific anatomical region capabilities not acquired incidentally from descriptive supervision alone.

The cross-complexity results (Tables~\ref{tab:cross_L2}-\ref{tab:cross_L3}) support this view. For L2, training InternVL2-8B on L1 alone raises Subset Accuracy from 0.1150 (zero-shot) to 0.1695 and Jaccard from 0.2509 to 0.2941 (Table~\ref{tab:cross_L2}). For L3, error falls monotonically with added lower-tier supervision: MAE decreases from 2.55 (zero-shot) to 2.52 (L1) to 2.50 (L2), reaching 2.46 when L1 and L2 are combined, with RMSE following the same trend ($4.71 \rightarrow 4.60$, Table~\ref{tab:cross_L3}).

Several limitations apply to our study. The radiographs come from a single institution and two acquisition devices (Planmeca ProMax 2D and Sirona Orthophos XG), so cross-scanner and cross-population generalization remain untested. Annotations rest on a single licensed dentist, and we do not report inter-annotator agreement, so label noise cannot be quantified. The six-region localization scheme (Fig. ~\ref{fig:dental_xray1}) is robust to missing or partially visible teeth but is coarser than tooth level indexing, trading spatial precision for annotation reliability. Within these constraints, DentiAsk is intended as a controlled testbed for measuring whether multimodal medical models can progress from recognition toward clinically grounded, query-conditioned reasoning.

\section{Conclusion} 
We introduced DentiAsk, a clinically grounded visual question answering benchmark for panoramic dental radiograph interpretation, comprising 1,000 high-resolution radiographs and 10,000 Human-validated question-answer pairs across periapical radiolucency, impacted teeth, and dental caries. Unlike dental datasets centered on segmentation or classification, DentiAsk unifies descriptive recognition (including impacted-teeth subtype classification), region-level localization, and numerical counting as separately scored tasks in a single evaluation framework, allowing a model to be tested on all of these capabilities over the same image. Benchmarking 10 vision-language models exposes a clear perception-reasoning gap: recognition is comparatively more tractable, while localization and counting, particularly across co-occurring pathologies, remain difficult and are not closed by model scale or by transfer from simpler tasks. We release DentiAsk and its evaluation protocol as a step toward multimodal models that interpret dental radiographs the way clinicians do, rather than merely recognizing that a finding is present.

\subsection*{Author contribution}
D. Jha conceived the study, developed the central research idea, designed the overall study framework, and guided dataset development and experimental design.  P. Pawar provided the panoramic dental radiograph dataset, and O. K. Susladkar coordinated with Dr. P. Pawar to facilitate the transfer and donation of the dataset to Dr. Jha’s research group. G. Reynolds contributed expert clinical annotation of the dataset. R. Paudel, A. Ghimire, A. Dhakal, S. Adhikari, and D. Jha contributed to dataset annotation, curation, verification, and quality control. T. K. Dutta performed the computational experiments. D. Jha led the writing and organization of the manuscript, with  contributions from T. K. Dutta and R. Paudel. All authors reviewed and edited the manuscript and approved the final version of the work.

\small
\bibliographystyle{unsrtnat}
\bibliography{neurips_2026}

\end{document}